\documentclass[preprint,12pt]{elsarticle}

\usepackage{amssymb}
\usepackage{subfigure}
\usepackage{vmargin}
\usepackage{fancyheadings}
\usepackage{graphicx}
 \usepackage{multirow}
 \usepackage{algorithm}
 \usepackage{algpseudocode}
\usepackage{amsmath,amssymb,amsthm,enumerate,bbm,color,verbatim,setspace}
\journal{Journal of Computational Physics}

\begin{document}

\begin{frontmatter}



\title{Highly parallel algorithm for the Ising ground state searching problem}


\author{ A. Yavorsky$^{1}$, L.A. Markovich$^{1,3,4*}$, E.A. Polyakov$^{1}$, A.N. Rubtsov$^{1,2}$}

\address{$^1$Russian Quantum Center, 100 Novaya St., Skolkovo, Moscow 143025, Russia, \\
$^2$M.V. Lomonosov Moscow State University, 119991 Moscow, Russia,\\
$^3$Institute for information transmission problems, Moscow,\\ Bolshoy Karetny per. 19, build. 1, Moscow 127051, Russia\\
$^4$V. A. Trapeznikov Institute of Control Sciences, Moscow,\\Profsoyuznaya 65, 117997 Moscow, Russia\\
$^*$Corresponding author e-mail: kimo1@mail.ru}

\begin{abstract}  
Finding an energy minimum in the Ising model is an exemplar objective, associated with many combinatorial optimization problems, that is computationally hard in general, but occurs in all areas of modern science. There are several numerical methods, providing solution for the medium size Ising spin systems.
%Among them, simulated annealing (SA) is considered as a good  tool for complex nonlinear optimization problems. 
%The method is widely applied to a variety of fields like physics, computer science, biology and economics.
%However, the major disadvantages of SA are that it is extremely slow and  hard to parallelise due to its sequential nature.
% Another approach is a Mean-Field Annealing (MFA) algorithm and its variations that are fast, but not accurate enough.
However, they are either computationally slow and badly parallelized, or do not give sufficiently good results for the large systems.
%This constantly motivates the attempts to develop novel algorithms, in order to obtain significant speed-ups and quality improvement. %for some specific tasks.
In this paper, we present a highly parallel algorithm, called Mean-field Annealing from a Random State (MARS), incorporating the best features
of the classical  simulated annealing (SA) and Mean-Field Annealing (MFA) methods. The algorithm is based on the mean-field descent from a randomly selected configuration and temperature. Since a single run requires little computational effort,  the effectiveness can be achieved  by massive parallelization.
MARS shows excellent performance both on the large Ising spin systems and on the set of
exemplary maximum cut benchmark instances in terms of both solution quality and computational
time. 

%Out of the 71 benchmark instances, BLS is capable of finding new improved results in 34 cases and
%attaining the previous best-known result for 35 instances, within computing times ranging from less
%than 1 s to 5.6 h for the largest instance with 20,000 vertices.
\end{abstract}


\end{frontmatter}


\section{Introduction}

\par In modern literature there are more and more examples when mathematical challenges have been solved using the solution of physical problems.
Thus, one of the most important tasks in combinatorial optimization - finding the extremum, is closely related to 
 the minimum energy estimation in the Ising model, proposed by Edwards and Anderson (EA)  \cite{Edwards} over forty years ago.
In physics the spin glass Ising model is  characterized by a complex energy landscape which possess many ultra deep local minima (the valleys).  This leads to the existence of a critical temperature $T_c$, below which the spins are frozen in random orientations (the so-called spin-glass transition). There is a combinatorially large number of such frozen spin configurations. Being caught in one configuration below $T_c$, the spin glass will never be able to escape to any other configuration. Experimentally this is observed as a peculiar behaviour of magnetization below $T_c$. Finding the ground states of the spin glasses is important, since this provides knowledge about properties of the low-temperature spin glass phase. For example, in  \cite{PhysRevLett.88.077201,PhysRevB.66.224401,PhysRevE.59.84} the stiffness exponent of the spin glass is computed, using the ground state energy. Besides its relevance in condensed matter, solid state and statistical physics, the spin glass theory is applied in such disciplines as  machine learning and neural networks \cite{Gardner_1988}. The Ising model lies at the heart of the Hopfield neural network model and the concept of the aforementioned energy valleys is used to analyse its memory capacity. Moreover, the spin glasess are used in many fields of computer science \cite{doi:10.1142/0271}, theoretical biology \cite{doi:10.1142/9789814415743_fmatter}, econophysics, information processing \cite{Nishimori}, mathematics \cite{stein2013spin}. 
Thus, the simple mathematical description of the Ising spins model leads to the fact that it became a benchmark in the complexity theory. It turned out, that any combinatorial $NP$-hard problem can be reduced to the problem of finding the ground state of the Ising model.
For example, it is equvivalent to such combinatorial optimisation problems as the travelling salesman (TSP) \cite{PAPADIMITRIOU1977237} or the maximum cut (MAX-CUT) \cite{Karp1972}.  TSP finds an enormous amount of applications such as data association, vehicle routing \cite{MARINAKIS2010463}, data transmission in computer networks \cite{286889}, scheduling, drilling of printed circuits boards, analysis of structure of crystals, clustering of data arrays, imaging processing and pattern recognition \cite{827167}.  The problem of cluster analysis,
 where a set of data points is  partitioned into sets of related observations, can be  modelled as MAX-CUT. 
\par 
Finding the ground state in the Ising model is a difficult task. 
For the general model, where all the spins interact with each other, called Sherrington-Kirkpatrick (SK) model \cite{Sherrington_Kirkpatrick}, 
a consistent conjecture for the asymptotic free energy per spin was proposed by Giorgio Parisi in 1982 (see \cite{Parisi1,Parisi2}).
The free energy function was obtained by Onsager (1944) (see \cite{PhysRev.65.117}). An infinite two-dimensional
grid was obtained in the ferromagnetic case,  when all interactions are equal to one and without
a magnetic field. Later the results for other planar two-dimensional lattices were obtained, but no general 
conclusions for the three-dimensional lattice or the  two-dimensional lattice within a magnetic field were provided in literature. 
Thus, finding the exact solution to the SK problem remains an open problem. However, one can use the numerical methods  to find the ground state quite precisely.
Since the total number of states for a structure with $N$ spins is $2^N$, as
soon as $N$ exceeds $30-40$, it is impossible, from a computational point of view,
to find a ground state by brute force, i.e., by enumerating all possible states
and computing the energy for each of them. Thus, the possibility of constructing an algorithm that, for any choice of the spin interactions and any magnetic field can find the ground state in a number of elementary operations, which are bounded by a polynomial function of $N$, is a central problem in computational mathematics.
If it is possible, the problem is polynomially solvable.
However, finding the ground state of the classical spin glass is $NP$-hard \cite{Barahona_1982} and finding the ground state for the quantum system with a local Hamiltonian is QMA hard \footnote{The class QMA  stands for Quantum Merlin Arthur. It is the quantum analog of the nonprobabilistic complexity class $NP$ or the probabilistic complexity class $MA$. 
It is the set of decision problems, where if the answer is YES, one can find a polynomial-size quantum proof (a quantum state) that convinces a polynomial-time quantum verifier of the fact with high probability. If the answer is NO, the verifier rejectes every polynomial-size quantum state  with high probability.}  \cite{doi:10.1137/S0097539704445226}. $NP$ - hard problems are reducible to each other by polynomial transformations. That means that ones a polynomial time algorithm is available for one member of the $NP$ - hard problems, all the $NP$ - problems are automatically solved. 
Despite no such method is known, the attempts to create both  algorithmic  and digital solutions  are undertaken in modern science.
 \par Various methods, based on deterministic or probabilistic  heuristics, have been developed to solve $NP$ - hard combinatorial optimization problems. One can name branch-and-bound \cite{Finke}, branch-and-cut \cite{DeSimone1996,DeSimone1995}, particle swarm optimization \cite{MARINAKIS2010463,MARINAKIS20101446}, tabu search \cite{doi:10.1002/mde.4090110512}, ant colony optimization \cite{GHAFURIAN20111256}, genetic algorithms \cite{867862}), self-organizing maps \cite{BAI20061082}, elastic net \cite{doi:10.1162/neco.1989.1.3.348}, Lagrangean relaxation \cite{ZAMANI201082}, etc. 
Likewise, a special purpose hardware that can solve $NP$-hard problems
more efficiently than the classical computers, is an active area of research. As an alternative to the
current von Neumann computer-based methods,  a neural network, realized with analog electronic
circuits is presened in \cite{Hopfield625,CREPUT20091250}.  
 Other interesting approaches are the molecular computing \cite{Adleman1021} and  the adiabatic quantum computation \cite{Farhi472}.
\par In 80's, the simulated annealing (SA) algorithm for the combinatorial optimization problems was introduced \cite{Kirkpatrick,GENG20113680}. This algorithm is inspired  by the thermal annealing procedure in metallurgy \cite{Kirkpatrick1983OptimizationBS,Brooke779}. It is still unsurpassed in its combination of clarity, simplicity, universality and reliability. Nevertheless, the abundance of ultra deep local minima in the Ising model necessitates an exponentially large simulation time in order to obtain a reasonable estimate of the global minimum.  This quickly make the simulation infeasible for large problem sizes. Another widely used approaches are the mean-field annealing, parallel-tempering Monte-Carlo \cite{PhysRevLett.57.2607}, population annealing Monte-Carlo \cite{PhysRevE.92.013303}, and others. SA-like algorithms are relatively easy to implement. Since SA  statistically provides an optimal solutions for many combinatorial problems (cf. \cite{INGBER199329}), it can be used as a bona fide method. However SA-like methods depend on many parameters  and are quite sensitive to the cooling scheduler.  The possible SA optimizations can be found, for example, in \cite{ISAKOV2015265}. An obvious approach to speed up this class of algorithms is parallelism. Unfortunately, by its nature, the method is hardly  parallelizable.  Despite to that, in literature one can find several attempts to construct such an algorithm. The clustering algorithm and  the genetic clustering algorithm \cite{RAM1996207} are the good examples. However, basically the parallelisation methods are provided for the simplest Hamiltonians, where the interaction occurs only between the nearest neighbours of the spin, selected at each MC step \cite{ALLWRIGHT1989335,144393}. 
 \par %The purpose of this paper is to propose a highly parallelizable algorithm, combining the best features of the methods known in the literature.
  Our approach is inspired by the recent developments of physical simulators. These are hardware  devices, based on various physical principles e.g. on the interaction between optical pulsed as in the coherent Ising machine (CIM) or network of non-equilibrium bose-einstein condensates \cite{PhysRevA.88.063853,Inagaki603,Yamamoto}. These devices are designed in such a way that the spin variables of the optimization problem are mapped on a physical continuous degrees of freedom of the device (optical quadrature or bose-einstein condensate phase).  The solution to the combinatorial optimization problem is obtained as a state of these degrees of freedom after manipulating the device  according to a certain protocol.
The latter machines are expensive and often designed for a very specific class of problems. Thus, a new interesting noisy mean-field annealing (NMFA) algorithm that emulates the operation of the CIM, is proposed in  \cite{NMFA}. It is shown that NMFA performs comparably to the CIM but runs roughly 20 times faster in absolute terms.
 Another efficient method of simulating the CIM on a classical computer, called the SimCIM, is introduced in \cite{Tiunov}. The algorithm outperforms both the CIM and the NMFA. One can conclude, that algorithmic ideas are frequently used in physical systems, and vice versa, modelling and analysis of the physical systems leads to new algorithmic ideas. The latter results give hope for creation of even faster and more accurate algorithms, inspired by the quantum machines and classical annealing method.
\par In this work we propose a new highly parallel method, called \textit{Mean-field Annealing from Random State (MARS)}, for solving the Ising ground state search problem, where the interaction occurs between all the neighbours. The algorithm combines the Mean field-like search with the special starting annealing temperature selection. On every simulation  the starting configuration and the maximum temperature of the descent  are randomly selected. The temperature is bounded by a given range and the descent is performed by solving the field equations, that is computationally fast operation. Moreover, each descent can be done separately and the algorithm is easy to  parallelize. Also MARS is not  sensitive to the cooling scheduler since only the boundary values of the random starting temperature are important. In fact, having a sufficiently powerful computer cluster with a large number of cores, one can instantly obtain the necessary statistics and find a solution to the optimization problem in much less time than any of the algorithms, known from literature for our best knowledge.
Despite  its simplicity, our algorithm shows excellent performance on the large Ising spin systems and on the set of known MAX-CUT instances in terms of both solution quality and computational time.
% Out of the benchmark instances, the proposed algorithm shown better results then best-known in literature for $?$ cases and reaches the previous best-known results for $?$ graphs for a little computational time.
 \par The paper is organized  as follows. In Sec.~\ref{sec:1} we  recall some basic ideas about the Ising spin glass model and its application in solving the MAX-CUT problem. In Sec.~\ref{sec:2} a brief overview on some commonly used algorithms for searching the Ising ground state is provided.
In Sec.~\ref{sec:3}    the new Mean-field from Random State approach for the Ising and MAX-CUT problems is presented. In Sec.~\ref{sec:4} the extensive computational results and comparisons with SA, NMFA and SimCIM methods are provided. 
%Moreover, we investigate the efficiency dependence of our method and SA, NMFA and SimCIM on the temperature scheduler.
 
 \section{Ising Spin Glass Model}\label{sec:1}
\par  
The modern theory of the spin glasses began with the work of Edwards and Anderson (EA) \cite{Edwards} who proposed the
simplest Hamiltonian that models the spin glasses, where only the nearest neighbours interacted.  However, this restriction does not occur in the real spin glass materials. 
 The infinite-ranged version of the EA Hamiltonian is proposed by Sherrington and Kirkpatrick (SK) \cite{Sherrington_Kirkpatrick}.
The system of $N$ spins is coupled by a pairwise interaction
\begin{eqnarray}
H_N(\sigma)=\sum\limits_{i}\sum\limits_{j	\neq i}J_{ij}\sigma_i\sigma_j+\sum\limits_ih_i\sigma_i, \label{2}
\end{eqnarray}
 where $J_{ij}$ are symmetric independent identically distributed random variables (iid rvs),  chosen from a Gaussian distribution with zero mean and variance one, $\sigma=(\sigma_1,\dots, \sigma_N)\in 	\Sigma_N=\{\pm1\}$ is the spin configuration. 
 The model contains the external field term $h_i$. For simplicity we will assume it zero, but all the results hold in the presence of the field with some minor modifications.
 \subsection{Ground state of Ising model: A physicist's perspective}
\par In  the SK model two mathematical problems arise. 
The first one is the study of the minimum energy configuration $\min_{\sigma\in \Sigma_N}{ H_N(\sigma)}$, called the ground state, and to  understand its behaviour in the thermodynamic limit $ N \rightarrow \infty$. 
%The second one is the calculation of the magnetic partition function $Z_N$. It can be reduced to the computation of the thermodynamic limit of the free energy 
%\begin{eqnarray}\label{20_1}
%F_N=-kT\log{Z_N}, \quad Z_N=\sum\limits_{\sigma\in \Sigma_N}\exp{(-H_N(\sigma)/kT)},
%\end{eqnarray}
%where $\beta=1/T>0$ is the inverse temperature parameter, $k$ is the Bolzmann constant.  The
%equilibrium magnetic properties, magnetisation, entropy, magnetic energy, specific
%heat and susceptibility, can all be obtained by differentiating the free energy with respect
%to magnetic field and temperature. 
\subsection{Ground state of Ising model: Combinatorial optimization perspective}
Most of the  combinatorial optimization problems can be reduced to the problem of finding the  spin configuration, corresponding to the ground state of the Ising model. In  this paper we consider  the maximum cut problem in a weighted graph (MAX-CUT), a classical problem in combinatorial optimization.
Let us have an undirected graph $G = (V,E)$, where $V=\{1,\dots,n\}$ is the set of vertices, $E$ is the set of edges and the matrix $J_{ij}$ of weights, associated with the edges $(i,j)\in E$. The essence of MAX-CUT is to  find a cut $(S,V 	\setminus S)$,  such that the sum of the weight of the edges with one endpoint in $S\in V$ and the other in $V	\setminus S$ is maximized over all possible cuts.
MAX-CUT is proofed to be a $NP$ - hard problem with applications in several fields,
including VLSI design and statistical physics (cf. \cite{doi:10.1287/opre.36.3.493}) since
it is equivalent to the Ising problem. The cut value can be written as
\begin{eqnarray}\label{3}
J(S,V	\setminus S)=\frac{1}{4}\sum\limits_{i,j\in V}J_{ij}-\frac{1}{4}\sum\limits_{i,j\in V}J_{ij}\sigma_i\sigma_j.
\end{eqnarray}
where the spin value $\sigma_i\in\{-1,1\}$, $\forall i\in V$ encodes which subset the $i$th node belongs to.
The cut value is maximized if the Ising energy is minimized.  The two-dimensional grid model, where 
only the nearest neighbours interact, no periodic boundary conditions and no
magnetic field, hold, is equivalent  to the problem of solving the MAX-CUT problem in a planar
graph. Thus, solving the physical problem, one can  obtain the solution to some mathematical tasks.
\section{Simulated Annealing methods} \label{sec:2}
\par
 In this section, we briefly outline the numerical methods used in the literature to solve the mentioned class of problems.
We start from \textit{Simulated annealing}, that  is a Monte-Carlo method of Metropolis et al. \cite{doi:10.1063/1.1699114} with a temperature schedule  \cite{Kirkpatrick1983OptimizationBS}, that can be modelled mathematically, using the finite Markov chains theory. 
The Markov chain is a sequence of trials, where the probability of the outcome of a given trial depends only on the outcome of the previous trial. 
In SA, a trial corresponds to a candidate solution which is to be optimized. The set of outcomes is given by a finite set of neighbouring states. Each move depends only on the results of the previous step of the algorithm. 
The SA algorithm is based on the following steps:
\begin{itemize}
\item Choose a random configuration $\sigma_i$, select the initial system temperature, and specify the
cooling (i.e. annealing) schedule. Evaluate energy $ E(\sigma_i)$.
\item  Perturb $\sigma_i$ to obtain a neighbouring trial vector $\sigma_{i+1}$. Evaluate $ E(\sigma_{i+1})$.
\item If $ E(\sigma_{i+1})< E(\sigma_i)$, $ E(\sigma_{i+1})$ is the new current solution. Otherwise, the configuration  $\sigma_{i+1}$ is
accepted as the new current configuration with a probability $\exp{[-(E(\sigma_{i+1})- E(\sigma_i))/T]}$.
\item   Reduce the system temperature according to the cooling schedule.
\end{itemize}
The SA presents an optimization technique that has advantages and disadvantages, compared to other global optimization methods, such as \textit{genetic algorithms, tabu search} and \textit{neural networks}.
It is extremely sensitive to the temperature schedule, namely it depends on the initial and final temperature, as well as on the temperature reduction law. A stopping criterion is chosen, which can be the maximum number of steps, the target minimum temperature or the freezing of configuration.
In other words, one performs the Monte Carlo simulation of the Ising spin system, starting at the high temperature. Using some cooling schedule, the temperature is slowly decreased during the simulation and  the configuration of the system falls into a local minima. If the schedule and the starting temperature  are selected 
correctly, with multiple repetitions, the global minimum can be estimated. One must point out, that the true strength of the SA method is that it statistically provides a true global optimum. %However, this does not mean that it is more effective than any other algorithm that can find a global optimum.
Most of SA-like algorithms, like \textit{simulated quenching} (SQ),  \textit{fast annealing} (FA) or  \textit{adaptive simulated annealing} (ASA) \cite{INGBER1996}  differ from each other by the annealing schedule.  BA and FA have one annealing schedule for $N$ distributions, which sample infinite ranges.  We will not focus on the latter methods, an interested reader can see the detailed review in 
\cite{INGBER199329}. 
\par Another formulation of the SA method, based on the mean field theory, is given in \cite{Bilbro:1988:OMF:2969735.2969746}. 
In \textit{Mean Field annealing} (MFA) on every MC step  a randomly selected discrete spin  $\sigma_i$ is replaced  by a continuous spin average $\langle \sigma\rangle$. Unlike SA, which solves the exact statistical physics problem, MFA is obtained from SA as the mean-field approximation.
The MFA algorithm is based on the following steps:
\begin{itemize}
\item Initialize the spin averages $\langle \sigma_i\rangle=1/2+\delta$, where $\delta$ is an added Gaussian noise.
\item Perform the following relaxation step until a fixed point is found:
\begin{enumerate}
\item Randomly select $\langle \sigma_i\rangle$.
\item Compute the mean field 
$\Phi_i=h_i+2\sum J_{ij}\langle \sigma_j\rangle$.
\item Compute the new spin average as $\langle \sigma_i\rangle=(1+\exp{(\Phi_i/T)})^{-1}$.
\end{enumerate}
\item Decrease temperature and return to the previous step.
\end{itemize}
The temperature is decreased according to the selected scheduling regime. Repeating the procedure, one can find the optimum solution.
The algorithm works fast, but for the large spin systems does not give satisfactory results.
\par A promising MFA-like method, called the \textit{Noisy Mean Field annealing} (NMFA), that  is a mathematical model of the \textit{Coherent Ising Machine} (CIM), is introduced in \cite{NMFA}.
In contrast to the standard MFA, the  NMFA adds a Gaussian noise $N(0,\sigma)$ to the normalized mean-field terms, namely
 \begin{eqnarray*}\Phi_i=(h_i+\sum_j J_{ij}\sigma_j)/ \sqrt{h_i^2+\sum_j J_{ij}^2}+N(0,\sigma).\end{eqnarray*} 
At each iteration (step) the corresponding thermal spin averages are  \begin{eqnarray*}\hat{\sigma}_i=-\tanh{(\Phi_i/T)}.\end{eqnarray*} After that, the convex combination is taken as the new spin value $\alpha \hat{\sigma}_i+(1-\alpha)\sigma_i$ and the temperature is decreased according to the selected scheduler.
Authors claim that the NMFA algorithm fully emulates CIM and performs similarly, solving the MAX-CUT problem. However, since the algorithm runs on a classic computer, the cost of its work is minimal.
\par Finally, we would like to mention the new algorithm, called SimCIM \cite{Tiunov},  that also successfully emulates CIM. The discrete spin variables $\sigma_j$ are replaced with continuous variable $X_j$. Next, the gradient of the rewritten Hamiltonians energy, namely $F=-\bigtriangledown_j H=1/2 \sum_jJ_{ij}\sigma_j$, is found. Authors affirm that SimCIM outperforms NMFA and CIM, that is demonstrated on several examples.
\subsection{Parallelisation of SA}
Since SA is based on a Markov chain sequence it is not straighforward how to run it on parallel processors. However, there have been many attempts to develop the parallel versions of the algorithm.  Basically the parallelisation methods are provided for EA Hamiltonians, where the interaction occurs between the nearest neighbours of the spin, selected at each MC step \cite{ALLWRIGHT1989335,144393}.
In literature, one can find two different approaches to the parallelization of SA:
the single-trial parallelism and the multiple-trial parallelism.
In the  single-trial version the calculations to evaluate a single trial are divided between
several processors. It is clear, that this
strategy and the possible speed-up are problem dependent. In multiple-trial parallelism, all the trials are evaluated in parallel. The latter approach is  also problem-dependent. In both cases one needs to divide the problem into subproblems and subsequently distribute
them  among the nodes or processors \cite{EGLESE1990271}.  Since the division of the problem into subproblems depends largely on the characteristics of the problem, any of the provided "parallel" SA algorithms cannot be general in nature. Also, it is possible that one node gets
into other nodes' search space, which  terms a collision. In the case of EA Hamiltonian, this is a rare event. On the other hand, in SK case, where all the spins interact with each other, this fact plays crucial role and have to be treated separately. 
An interested reader can find an overview on some parallel realizations of SA method like \textit{Clustering algorithm} or \textit{Genetic Clustering algorithm}  in \cite{RAM1996207,MAHFOUD19951}. 
\section{Mean-field Annealing from Random State}\label{sec:3}
To overcome the basic problem of SA-methods, its low computational speed, we introduce a highly parallel algorithm, called \textit{Mean-field Annealing from Random State (MARS)}. In contrast to the previously mentioned algorithms, it is not problem dependent and can be successfully applied to the SK model with a big amount of interacting spins. 
In MARS the choice of the initial temperature value, selected from a given range, plays the main role. Selecting the maximum temperature of the scheduler regime, the temperature is decreased until a solution is found. Making a parallel series of such descents, a sample of the intended solutions is formed from which the best solution is chosen.
\par Pseudocode is given below and source code is provided in supplementary  materials ($https://github.com/Yxbcvn410/Sherrington-Kirkpatrick$). 
Our algorithm stores spins $s_i$ as continuous values between $-1$ and $1$.
The mean-field term $\Phi_i$ is calculated for every spin and converted into a spin value $\hat{s}_i$, using the Boltzmann expectation at the current temperature $T_t$, namely $\hat{s}_i =- \tanh{(\Phi_i / T_t)}$. $T_t$ is a local variable, where the   systems temperature $t$ is stored. While processing the algorithm, the temperature decreases with the step $C_{step}$ to zero. The difference between the spin $s_i$ and the trial spin $\hat{s}_i$ is compared with the 
 stabilization parameter $d$, that is bounded by $d_{min}$. If $|\hat{s}_i - s_i| > d$, holds, the parameter $d$ is replaced by $|\hat{s}_i - s_i|$ and the spin $s_i$ with $\hat{s}_i$. The procedure is repeated, according to the temperature scheduler $t(T_{min},T_{max},T_{step})$. $T_{min}$ and  $T_{max}$ are the
  boundary values of the temperature range in which the analysed points are located, $T_{step}$ is the temperature step. 
The number of points (MC steps) is expressed as $(T_{max}-T_{min})/T_{step}$. 
\begin{algorithm}  
\caption{MARS generates a set of the Ising spins $s_i$, for the given Ising problem $(h_i$,$J_{ij})$.}
\begin{algorithmic}[1]
    \For{$t = T_{min}$ to $T_{max}$}
        \For{$i = 0$ to $N$}
         \State $s_i:=rand(-1, 1)$
           \EndFor
    \State  $T_t := t$
 \While{$T_t > 0$} 
       \State $T_t =T_t- C_{step}$;
       \Repeat
          \State $d = 0$
            \For{$i = 0$ to $N$}
               \State  $\Phi_i = \sum_{j}J_{ij}s_j$
               \State  $\hat{s}_i =- \tanh{(\Phi_i / T_t)}$
           
                \If{ $|\hat{s}_i - s_i| > d$}
                 \State  $d = |\hat{s}_i - s_i|$ 
   \EndIf 
     \EndFor
   \State  $s_i = \hat{s}_i;$   
      \Until{$d > d_{min}$} 
     \EndWhile
    \EndFor

\end{algorithmic}
\end{algorithm}
\par We implement this procedure on the \textit{Nvidia Tesla $V100$} video processor for the parallel calculation of the algorithm. In CUDA, all threads are combined into thread blocks, while all the blocks form a structure, called grid. There are usually $1024$ threads in one block (depending on hardware), and these blocks are easily synchronized so, that no thread in block proceeds from some point until all other blocks reach this point.
In the beginning of the algorithm we load matrix data to the video processor's memory and allocate memory for other variables. Then we run a substantial amount of blocks in parallel, where each block implements a single MC pass and has its own part of the memory with $s_i$ spin values. After all blocks finish working,  the data  is written to files $data\_hamiltonian.txt$ and $data\_maxcut.txt$. The new spin values are loaded and the new block set is launched.
While the program works, the block set working times are reflected in $log.txt$ file, and all the data about the best passes, including the best and the mean Hamiltonian and maximum cut, spin values, the best run quantity and the starting temperatures are stored in the file $spins.txt$.
The elementary operation of writing $-\tanh{(\Phi_i / T)}$ is parallelized as follows:
\begin{itemize}
\item Calculate and assign $s'_j = s_j  J_{ij}$ values. To this end the memory is allocated in the very beginning. This operation is easily distributed to the different streams. In time this piece is $O(1)$.
\item Sum up all the $s'_j$ values. Firstly, we assign $s'_{2i} = s'_{2i} + s'_{2i+1}$. This operation is also easily distributed to the different streams and therefore is $O(1)$ in time. Next $s'_{4i} = s'_{4i} + s'_{4i+2}$ is assigned, then $s'_{8i} = s'_{8i} + s'_{8i+4}$, etc. In $\log{2(N)}$ operations the sum of all $s'_i$ values will  be stored in $s'_0$, so the whole step will take $O(log(N))$ in time.
\item Assign $s_i = s'_0$. This operation is elementary and cannot be parallelized.
\end{itemize}
It is important to synchronize the flows between the steps of the algorithm and between the operations of the second step.

%We implement this  procedure on the GeForce 1080 graphical processor for parallel calculation of the algorithm.
%In CUDA, streams are combined into blocks. There are  1024 streams in the block, 
%that can be easily synchronized. We run a decent amount of blocks in parallel, where
%each block implements one MC pass. After they finish working, the data is
% processed  and the next set of blocks is launched. 
% The running time of the block set (in CUDA terminology - the grid) 
% is reflected in the log.txt file. The number of best passes and the total number of passes is tracked in the file spins.txt.
%\par The elementary operation - writing $-\tanh{(\Phi_i / T)}$ to $s_i$ is  parallelized 
%as follows.
%\begin{itemize}
%\item To calculate  $s'_j=s_jJ_{ij}$ the memory is
% allocated.  This operation is easily distributed to the different streams.
%  In time this piece is $O(1)$.
%\item The sum over all $s'_j$ is done as follows. First, for all $i$ the 
%$s'_{2i}= s'_{2i} + s'_{2i+1}$ is calculated. Next 
%$s'_{4i}= s'_{4i} + s'_{4i+2}$, $s'_{8i}= s'_{8i} + s'_{8i+4}$, etc.
%Through $\log_2(N)$ operations  the sum of all elements will be stored in $s'_0$. 
% One operation takes $O(1)$ time, so this step will take $O(\log{N})$ time.
% \item The operation $s_i = s'_0$ can be straightforwardly parallelized. 
%\end{itemize}
%It is important to synchronize the flows between the steps of 
%the algorithm and between the operations of the second step, i.e., no one thread proceeds to perform the next operation until all the others have completed the previous one. 
\section{Simulation results}\label{sec:4}
In this section, we report extensive computational results of our approach and show comparison with SA and  recently proposed  NMFA and SimCIM algorithms. We conduct our experiments on the variety of the Ising spin problems and on a set of  graphs from $G$ - set that has been widely used to evaluate MAX-CUT algorithms.
$G$ - set include toroidal, planar and random graphs, with number of vertices ranging from 
$800$ to $20000$ and edge weights of values $\pm 1,0$.
\subsection{Parameter settings and comparison criteria}
We use $C_{step}= 1$ and  $d_{min}=0.0001$ for all experiments in this paper.
The values of $T_{min}$, $T_{max}$ are different for each $J$ matrix and G-graph. The parameters are selected by performing a preliminary experiment on a selection of one size graphs and matrices. In Sec.~\ref{sec:4_1} we provide a parameter sensitivity analysis and justify the setting of parameters that is used to obtain the reported results. 
\par The assessment of our algorithm performance is based on comparisons against the best known results, reported in literature \cite{BENLIC20131162} and against three state of art methods. We show the best objective value, average  objective value and computational time.
It is obvious, that the comparison with the data from literature is not fully fair, since the computing hardware and programming languages are different. Thus, we mainly look on the best known result value (Tables~\ref{tab:3},\ref{tab:4}). That is why we did the comparison with several known algorithms like SA, NMFA and SimCIM fairly, i.e all the methods are done on one programming languages  and launched under one computing hardware. 
\subsection{Ground state of Sherrington-Kirkpatrick spin-glass}
\begin{figure}[ht!]  
\centering \subfigure[]{
\includegraphics[width=0.4\linewidth]{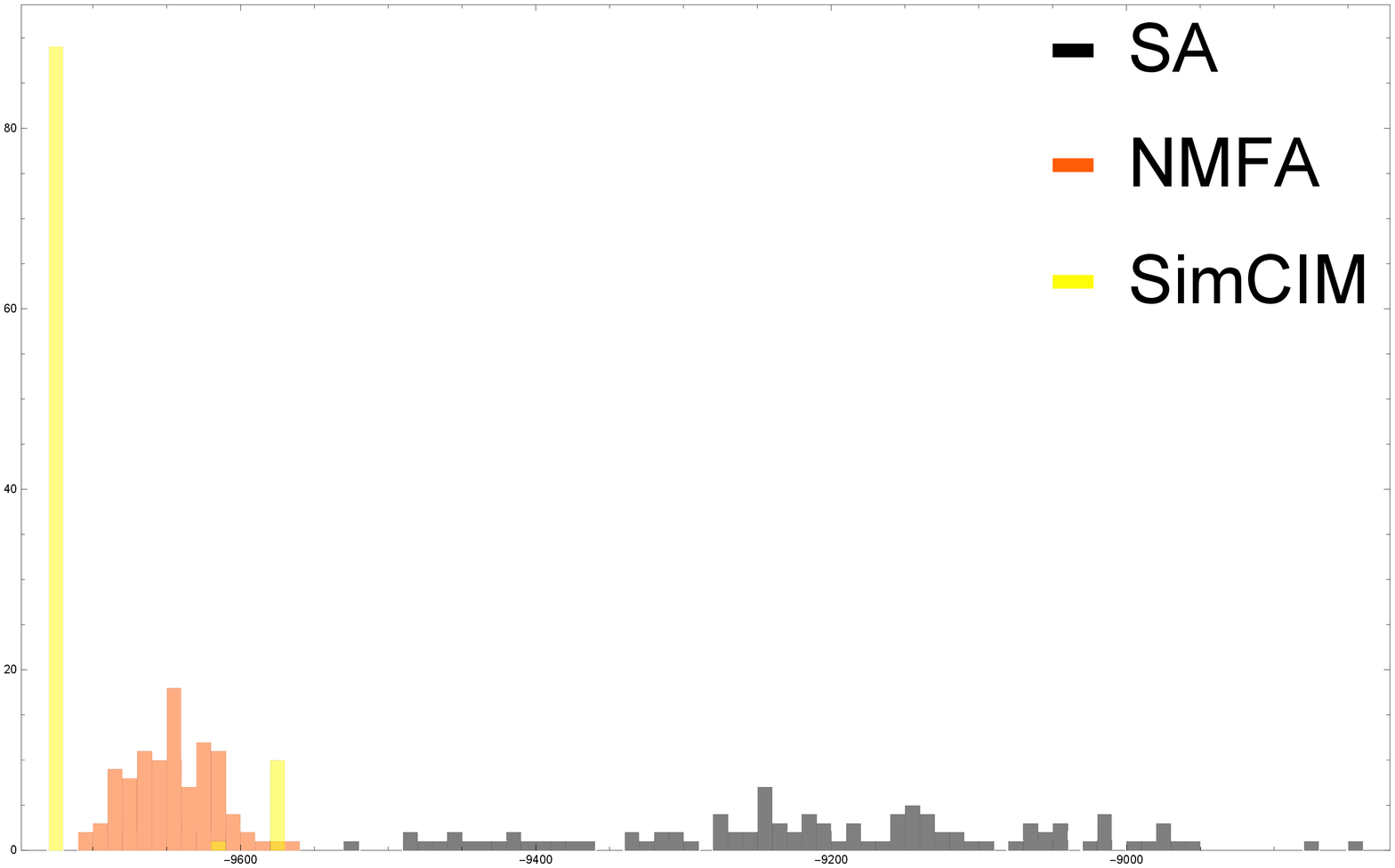} \label{fig:1} }  
\hspace{4ex}
\subfigure[]{
\includegraphics[width=0.4\linewidth]{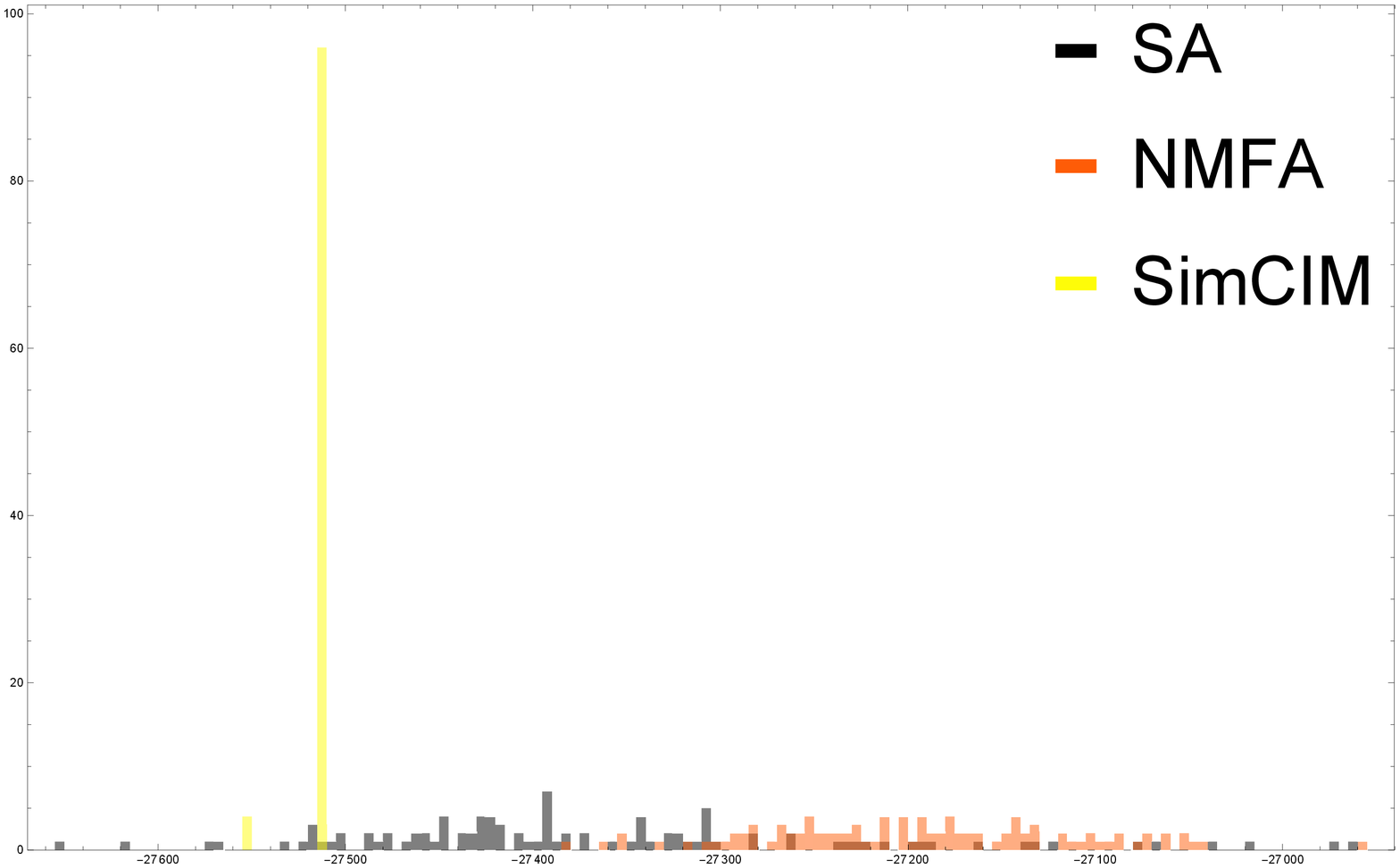} \label{fig:2} }
\hspace{4ex}
\subfigure[]{ \includegraphics[width=0.4\linewidth]{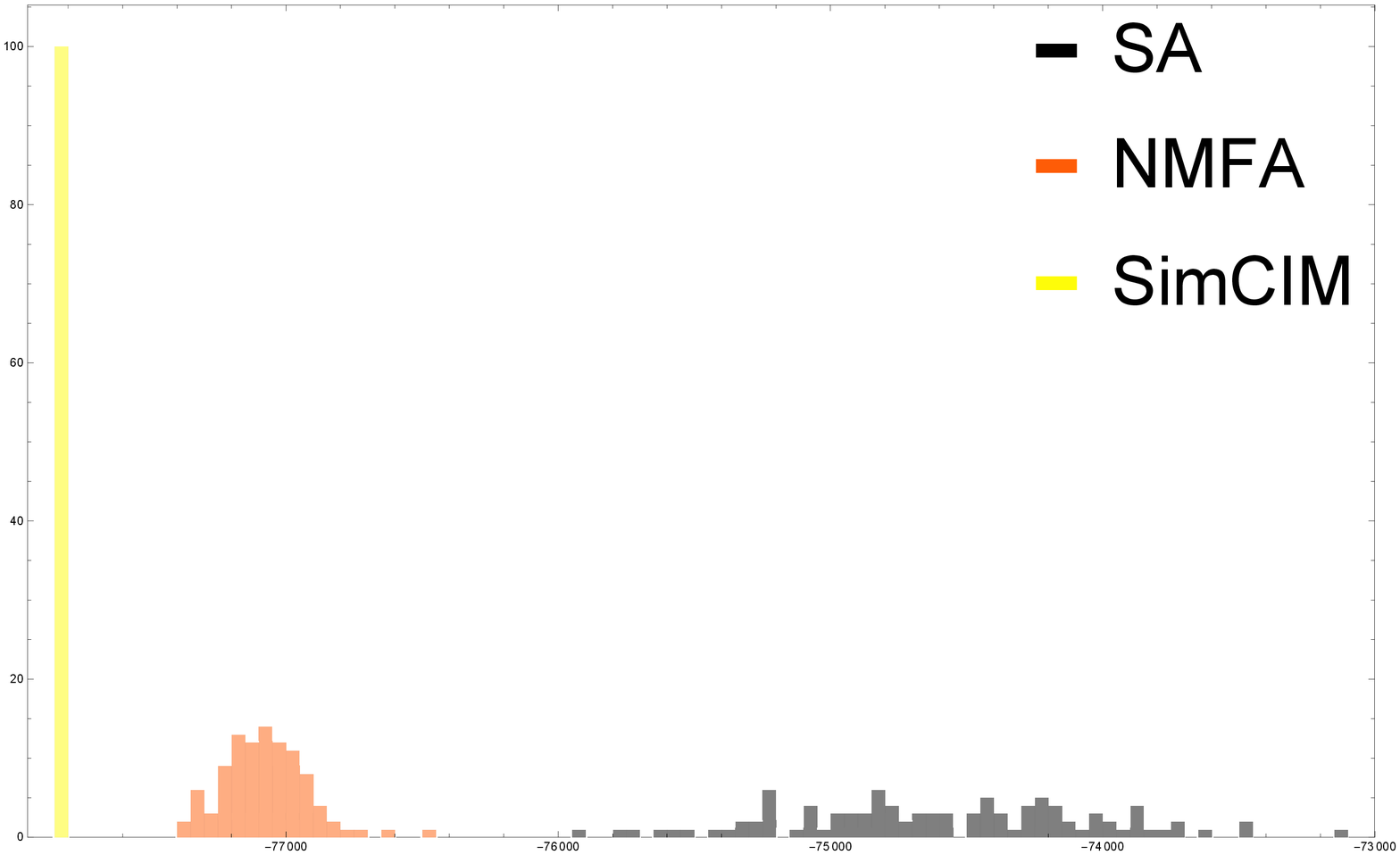} \label{fig:3} }  
\caption{Energy histograms  for the four Ising problem $(h_i=0,J_N)$: \subref{fig:1} $N=500$; \subref{fig:2} $N=1000$; \subref{fig:3} $N=2000$.} \label{fig:threeDMcases}
\end{figure}

As a first benchmarking problem, we select the SK spin-glass model on a fully - connected  graph, where the couplings $J_{ij}$ are iid Gaussian rvs. For each problem size $500\le N\le 2000$, we perform a set of simulations. As a performance metric we consider 
minimal energy and the success probability $P$, defined as the fraction of simulations on the same instance that return the ground state energy.
To get an idea of how long it takes each solver to draw a sample, we compare the mean runtimes of SA, NMFA and SimCIM
against our algorithm.
% Another metric is the multiplication of the expected number of independent  runs to solve a problem with $99\%$ probablity with  the time of a single run  $T_a$, namely
% \begin{eqnarray*}T_s=T_a(\log{(0.01)}/\log{(1-P)}).\end{eqnarray*}
\begin{table}[h]
\caption{\label{tab:1} Best and mean performance on $500,1000,2000$-spin Ising problem.}%(absolute value)
\begin{center}
\begin{tabular}{|c|c|c|c|c|c|c|c|}
\hline
\multirow{3}{*}{Algorithm}  & \multicolumn{6}{c|}{$J_N$}                                                                                               \\ \cline{2-7} 
                        & \multicolumn{2}{c|}{$500$} & \multicolumn{2}{c|}{$1000$  } & \multicolumn{2}{c|}{$2000$}  \\ \cline{2-7} 
                        & Best (Mean)    & Time(m)   & Best (Mean)     & Time(m)    & Best (Mean)      & Time(m)              \\ \hline
SA           &           $-4761.01$      & $0.090984$      &        $-13827.$    &  $9.18911$  &    $-37953.1$          &         $ 10.7057$          \\
          &           $(-4601.31)$      &     &        $ (-13681.6)$    &   &    $ (-37299.7)$          &                           \\ \hline
 NMFA    &     $-4850.72$       &   $0.343911$     &    $-13691.4$              & $2.30639$       &         $-38690.6 $        &   $24.3457$           \\ 
     &        $( -4822.98)$        &        &           $( -13597.5)$        &        &      $(-38537.9)$           &                \\ \hline
SimCIM &       $-4861.14$        &    $0.792329$      &         $-13775.1$      &    $ 3.37266$       &     $-38903.5$            &  $ 23.6815$              \\
 &     $( -4853.33)$            &        &     $(-13757.5)$            &         &          $(-38903.)$          &                    \\ \hline
MARS &       $-4865.16 $        &    $0.0014$      &         $ -13826.9$      &    $0.032$       &     $-39091.1$          &  $1.74098$                 \\
 &     $(-4760.72)$            &        &     $(-13534.1)$            &         &          $$          &                    \\ \hline

\end{tabular}
\end{center}
\end{table}
The results are presented in  Table~\ref{tab:1}. The minimal energy and mean runtimes (in minutes) for the four algorithms is provided. All algorithms are  optimized to run  $\{3\times10^4,4\times10^4,10^5\}$ iterations for $N\in\{500,1000,2000\}$, respectively. The temperature scheduler and parameters for NMFA are taken from \cite{NMFA}, for SimCIM from \cite{Tiunov}. These algorithms run  $100$ times and the histograms of the ground state energies are constructed (see Fig.~\ref{fig:1}, \ref{fig:2} and \ref{fig:3}). 
In its turn the MARS  algorithm performs with the following parameters:
\begin{itemize}
\item for the $J$ matrix of $N=500$ size the minimum and the mean energies are $-4865.16$ $ (-4760.72)$. This result was computed, using $30000$ simulations in $0 h$ $ 42 m $ $21 s$ (the mean time for one simulation is $0.0847$ seconds).
The hit count is equal to  $1920$, that gives the success probability $P=0.064$.
The temperature bounds $T_{min}= 0$, $T_{max}=30$, hold.
\item for $N=1000$ the minimum and the mean energies are $ -13826.9$ $ (-13534.1)$. 
 This result was computed, using $10000$ simulations in $5 h$ $ 20 m$ $2 s$ (the mean for one simulation $0.48005$ seconds).
The hit count is equal to  $1$, that gives the success probability $P=2.5 \times 10^{-4}$.
The temperature bounds $T_{min}= 0$, $T_{max}=40$, hold.
 \item for $N=2000$ the minimum energy is $ -39091.1$. 
 This result was computed, using $100000$ simulations  in $2 d$ $ 21 h$ $ 38 m$ $ 15 s$ (the mean for one simulation is $2.50695$ seconds).
The hit count is equal to  $2$, that gives the success probability $P=2\times10^{-5}$.
The temperature bounds $T_{min}= 0$, $T_{max}=40$, hold, and  $100000$ steps are done in total.
 \end{itemize}
From Table~\ref{tab:1} one can conclude that on all the three matrices, our algorithm performs
better and many times faster then the other methods.   

\subsection{MAX-CUT}
Next, we study the performance of MARS  on MAX-CUT problem.
% We  run the simulations on sparse graphs $G22$ and $G39$ from the $G$-Set \cite{} and the fully connected graph $K2000$.
In Table~\ref{tab:2} we compare the proposed  method to SA, NMFA and SimCIM algorithms
for the graphs with  $|V|= 2000$ nodes that were generated by \textit{rudy}, a machine
independent graph generator written by G. Rinaldi (see for example in \cite{doi:10.1137/S1052623497328987}).
Graph $G_{22}$ is unweighted random graph with a density of $1\%$, $G_{39}$ is unweighted "almost" planar graph with random
edge weights from $\{-1, 1\}$. 
The third example is $K_{2000}$ that is a fully-connected complete graph. The SA,
NMFA and SimCIM are optimized to run $\{4\times10^4,15\times10^4,5\times10^4\}$ iterations for all  experiments.

% \usepackage{multirow}
\begin{table}[h]
\caption{\label{tab:2} Best and mean performance  on $2000$-node MAX-CUT instances.}%(absolute value)
\begin{center}
\begin{tabular}{|c|c|c|c|c|c|c|c|}
\hline
\multirow{3}{*}{Algorithm}  & \multicolumn{6}{c|}{$J_N$}                                                                                               \\ \cline{2-7} 
                        & \multicolumn{2}{c|}{$G_{22}$} & \multicolumn{2}{c|}{$G_{39}$  } & \multicolumn{2}{c|}{$K_{2000}$}  \\ \cline{2-7} 
                        & Best (Mean)    & Time(m)   & Best (Mean)     & Time(m)    & Best (Mean)      & Time(m)             \\ \hline
SA           &           $12855. $      & $3.93602$      &        $2347.$    &  $14.917$  &    $31397. $          &         $5.38188$          \\
          &           $(12751.4)$      &     &        $ (2299.36)$    &   &    $ (30823.3)$          &                           \\ \hline
 NMFA    &     $13344.$       &   $9.74812$     &    $2372.$              & $36.6222$       &         $33259.$        &   $ 13.4483$           \\ 
     &        $(13332.3)$        &        &           $(2371.98)$        &        &      $(33244.3)$           &                \\ \hline
SimCIM &       $13349. $        &    $0.046$      &         $2363. $      &    $0.0634$       &     $33278.$            &  $0.0432$              \\
 &     $ (13300.6)$            &        &     $(2358.)$            &         &          $(33209.)$          &                    \\ \hline
MARS  &       $13354$        &    $0.0234$      &        $ 2382 $      &    $0.03362 $       &     $33311$          &  $0.0855$                 \\
 &     $(13122.7)$            &        &     $(2310.96)$            &         &          $(32039.1)$          &                    \\ \hline

\end{tabular}
\end{center}
\end{table}
MARS  performs with the following parameters:
\begin{itemize}
 \item  for $G_{22}$ graph the maximum  and the mean cut are $13354 $ $ (13122.7)$.  This result was computed, using $40000$ simulations in $3 h$ $ 54 m$ $33 s$
 (the mean for one simulation is $0.351825$ seconds).
The hit count is equal to  $2$, that gives the success probability $P=5\times10^{-5}$.
The temperature bounds $T_{min}= 0$, $T_{max}=40$, hold.
\item for $G_{39}$ graph the maximum  and the mean cut are $2382 $ $ (2310.96)$. This result was computed, using $150000$ simulations in $1 d$ $4 h$ $1 m$ $43 s$ (the mean for one simulation is $0.67269$ seconds).
The hit count is equal to  $1$, that gives the success probability $P=6.66\times10^{-6}$.
The temperature bounds $T_{min}= 6$, $T_{max}=6.75$, hold.
  \item  for $K_{2000}$ graph the maximum  and the mean cut are $33311.$ $ (32039.1)$. This result was computed, using $50000$ simulations in $14 h$ $ 15 m$ $ 30 s$ (the mean for one simulation is $1.0266$ seconds).
The hit count is equal to  $1$, that gives the success probability $P=2\times 10^{-5}$.
The temperature bounds $T_{min}= 0$, $T_{max}=40$, hold.
 \end{itemize}
The results in Table~\ref{tab:2} show superior performance of MARS algorithm  with respect to all the three  algorithms as for the best cut as for the computational time. 
\subsection{Comparison with the current best-known solutions}
 We tested MARS performance on the $G$ - set of graphs collection, generated by the \textit{rudy} graph generator, that are the standard test set for graph optimization. The parameters for the grapgh generator are provided in \cite{doi:10.1137/S1052623497328987}.
% Graphs from $G_1$ to $G_5$ are unweighted random graphs with
%a density of $6\%$. Grapghs from $G_6$ to $G_{10}$ are the same graphs with random edge weights from $\{−1, 1\}$,
%$G_{11}$ to $G_{13}$ are toroidal grids with random edge
%weights from $\{−1, 1\}$. $G_{14}$ to $G_{17}$ are unweighted "almost"
%planar graphs, having as edge set the union of two  planar graphs. Finally graphs from $G_{18}$ to $G_{21}$ are the same almost planar graphs with random
%edge weights from $\{−1, 1\}$.

Tables~\ref{tab:3} and \ref{tab:4} provide the computational results of our method on the set of most commonly used MAX-CUT instances in comparison with the current best-known results (column $f_{prev}$ is provided in \cite{BENLIC20131162}). For our algorithm we report the best objective value $f_{best}$, the average objective value $f_{avg}$ and time $t(s)$, given in seconds. One can conclude that MARS reaches the best cuts stated in the literature or differs slightly from them for the graphs from $G_1$ to $G_{13}$ and from $G_{43}$ to $G_{50}$ that are unweighted random graphs and toroidal grids with random edge
weights. Other grapghs are from the "almost" planar family. On them, Mars shows slightly worse results.

\subsection{Parameter sensitivity analysis}\label{sec:4_1}
First we investigate the performance of SA, NMFA and SimCIM algorithms depending on the amount of MC steps. For all algorithms we do not change the temperature regimes. Thus, the three-step scheduler for  NMFA and SimCIM algorithms is simply stretched to a greater number of points, namely the descent is done slower.
We took the same matrices as for the experiment in Sec.~\ref{sec:4}. For three matrices  we performed $\{10^4,10^3, 10^2\}$ simulations with $\{10^3,10^4, 10^5\}$ MC steps. The results are presented in Tables~\ref{tab:3_1}-\ref{tab:5}.
 
One  can conclude that the performance of SA  is improved with the growth of the amount of MC steps, that is obvious, since for the small number of MC steps the algorithm cant operate correctly. With a large number of MC steps, the algorithm shows good results, but it works very slowly. For these reasons, it is not suitable for the matrices of higher dimension. Finally for  NMFA and SimCIM, there is no noticeable improvement, since both are strongly dependent on the temperature scheduler, which obviously must be rearranged for a larger number of MC steps. For $10^3$ MC steps, the algorithms work fast and show good results. However, for a larger number of MC steps they need temperature scheduler correction. The calculation speed of the algorithms is also  low.
\par In Table~\ref{tab:3_2} the performance of MARS depending on the amount of simulation blocks is shown. For any sample size $\{10^2, 10^3, 10^4\}$, our algorithm outperforms the results, given in Tables~\ref{tab:3_1}-\ref{tab:5}. In Table ~\ref{tab:3_3} the dependence of the performance of MARS from the temperature boundaries is shown. One can see, that the correct selection of $T_{min}$ and $T_{max}$ strongly affects the quality of the energy estimation and the exeptance probablity ratio.

\section{Discussion and conclusion}
Let us summarize our results. A strongly parallel algorithm MARS for the ground state searching problem  for the fully connected Ising spin system is proposed. The performance of the algorithm is compared with the classical simulated annealing  method and the two most recent algorithms NMFA and SimCIM, that emulate the operation of the CIM. On the example of the  big Ising spin systems it can be concluded ,that the proposed algorithm shows the best results for incomparably shorter time then SA, NMFA and SimCIM. The proposed algorithm also shows excellent results solving the MAX-CUT problem. Comparative results show that MARS is ahead of SA, NMFA and SimCIM algorithms both in terms of the best cut estimation and computational time. A study  of the parameter  dependence of the algorithms is conducted. MARS shows the best performance even on the small samples. Finally, a comparison with the best  results from literature, known for the MAX-CUT problem on the basis of the $G$ -set graph collection, is performed. For the random graphs and toroidal grids with random edge
weights MARS reaches the best cuts stated in the literature or differs slightly from them. For the the "almost" planar family our algorithm performs comparable to the best  known in the literature algorithms for a sufficiently small computational time.

Thus, having a multi core computer, Mars can speed up the solution of the  large dimension Ising ground state search problems and can be used as a powerful tool for solving many combinatorial optimization  problems.
%\section{acknowledgements}
%Markovich L.A.  was partly supported by a Foundation for Basic Research, grant 14-50-00150

\bibliographystyle{unsrt}
\bibliography{Bibliography}
\section{Appendix}
\begin{table}[h]
\caption{\label{tab:3} }%(absolute value)
\begin{center}
\begin{tabular}{|c|c|c|c|c|c|c|}
\hline
\multicolumn{1}{|c|}{Graph}  & \multicolumn{1}{c|}{$|V|$} & \multicolumn{1}{c|}{$f_{prev}$  } & \multicolumn{1}{c|}{$f_{best}$} & \multicolumn{1}{c|}{$f_{avg}$} & \multicolumn{1}{c|}{$t(s)$}   & \multicolumn{1}{c|}{$P$} \\ \cline{1-7} \hline
$G_1$ &           $800$      & $11624$      &        $11623$    &  $11539.9$  &    $0.0857$     & $0.11944$                      \\
$G_2$ &     $800$       &   $11620$     &    $\textbf{11620}$              & $11540.9$       &         $0.0894$    & $8\times 10^{-5}$           \\ 
$G_3$ &       $800$        &    $11622$      &         $11621$      &    $11541$       &     $0.0908$  & $0.07732$                 \\
$G_4$ &       $800$        &    $11646$      &        $\textbf{11646}$      &    $11563.8$       &     $0.0938$      & $0.01238$                \\
$G_5$ &       $800$        &    $11631$      &        $\textbf{11631}$      &    $11550.9$       &     $0.4441$   & $8\times 10^{-5}$                 \\
$G_6$          &           $800$      & $2178$      &        $2176$    &  $2093.54$  &    $0.0883$         & $58\times 10^{-5}$      \\
$G_7$   &     $800$       &   $2006$     &    $\textbf{2006}$              & $1925.04$       &         $0.0903$    & $14\times 10^{-5}$          \\ 
$G_8$ &       $800$        &    $2005$      &         $\textbf{2005}$      &    $1932.04$       &     $0.091$       & $0.00292$             \\
$G_9$ &       $800$        &    $2054$      &        $2053$      &    $1969.7$       &     $0.0872$        & $2\times 10^{-5}$             \\
$G_{10}$ &       $800$        &    $2000$      &        $\textbf{2000}$      &    $1919.99$       &     $0.0851$     & $14\times 10^{-5}$                      \\
$G_{11}$          &           $800$      & $564$      & $560$        &  $529.1$  &    $0.491$           & $11\times 10^{-5}$                 \\
$G_{12}$   &     $800$       &   $556$     &    $ 554$              & $523.405$       &         $0.447$ & $136\times 10^{-5}$             \\ 
$G_{13}$ &       $800$        &    $582$      &         $580$      &    $547.367$&  $0.4335$     &      $13\times 10^{-5}$                     \\
$G_{14}$ &       $800$        &    $3064$      &        $3054$      &    $3004.53$       &     $0.554$   &      $8\times 10^{-5}$                         \\
$G_{15}$ &       $800$        &    $3050$      &        $3040$      &    $2986.1$       &     $0.396$     &      $1\times 10^{-5}$                    \\
$G_{16}$ &           $800$      & $3052$      &        $3042$    &  $2982.48$  &    $0.31$                &      $1\times 10^{-5}$   \\
$G_{17}$   &     $800$       &   $3047$     &    $3033$              & $2985.39$       &         $0.387$  &      $1\times 10^{-5}$            \\ 
$G_{18}$ &       $800$        &    $992$      &         $986$      &    $900.934$       &     $0.2295$      &   $1\times 10^{-5}$                \\
$G_{19}$ &       $800$        &    $906$      &        $899$      &    $840.1$       &     $0.4255$    &           $33\times 10^{-5}$              \\
$G_{20}$ &       $800$        &    $941$      &        $940$      &    $844.882$       &     $0.2375$                & $1\times 10^{-5}$            \\
$G_{21}$ &       $800$        &    $931$      &        $920$      &    $889.691$       &   $0.034$  & $0.0001$                         \\
$G_{22}$ &       $2000$      & $13359$      &        $13352$    &  $13177.5$  &    $0.3198$ & $0.0001$                \\
$G_{23}$   &     $2000$      &   $13344$     &    $13336$              & $13177.3$       &   $0.2599$ &    $0.0001$              \\ 
$G_{24}$ &       $2000$        &    $13337$      &         $13325$      &    $13173.3$       &     $0.2872$  &$0.001$                     \\
$G_{25}$ &    $2000$       &    $13340$      &        $13326$      &    $13174.3$       &   $0.2626$ & $0.0002$                        \\
$G_{26}$ &    $2000$       &    $13328$      &        $13316$      &    $13167.3$       &     $0.2673$   & $0.0002$                       \\
$G_{27}$  &   $2000$      & $3341$      &        $3327$ &  $3175.12$  &    $0.2719$      & $0.0001$              \\
$G_{28}$   &   $2000$       &   $3298$     &    $3288$              & $3140.41$       &         $0.2333$       &       $0.0011$            \\ 
$G_{29}$ &       $2000$     &    $3405$      &         $3387$      &    $3235.86$       &     $0.2693$   &       $0.0002$                        \\
$G_{30}$ &       $2000$       &    $3412$      &        $3406$      &    $3246.59$       &     $0.2451$        &       $0.0002$                    \\
$G_{31}$ &      $2000$        &    $3309$      &        $3302$      &    $3148.86$       &     $0.279$           &       $0.0001$               \\
$G_{32}$ &       $2000$       &    $1410$      &        $1392$      &    $1325.79$       &     $0.1578$      &       $0.0001$                       \\
$G_{33}$ &       $2000$      & $1382$      &        $1368$    &  $1301.09$  &    $0.1797$    &       $0.0003$             \\
$G_{34}$ &     $2000$      &   $1384$     &    $1372$              & $1306.1$       &         $0.1673$  &       $0.0004$                  \\ 
$G_{35}$ &      $2000$       &    $7684$      &         $7629$      &    $7578.39$       &     $0.2589$    &       $0.0001$                       \\
$G_{36}$ &       $2000$    &    $7678$      &        $7624$      &    $7581.53$       &     $0.3476$      &       $0.0001$                       \\
$G_{37}$ &      $2000$      &    $7689$      &        $7632$      &    $7589.24$       &     $0.2842$   &       $0.0001$                        \\
$G_{38}$ &      $2000$     & $7687$      &        $7629$    &  $7589.28$  &    $0.3211$      &       $0.0001$             \\
$G_{39}$   &     $2000$       &   $2408$     &    $2382$              & $2309.95$       & $0.2857$ &       $0.0002$                 \\ 
$G_{40}$ &       $2000$       &    $2400$      &         $2349$      &    $2278.13$       &     $0.3497$ & $0.0003$                        \\
$G_{41}$ &      $2000$       &    $2405$      &        $2363$      &    $2241.28$       &     $2.825$     &   $1\times 10^{-5}$                        \\
$G_{42}$ &       $2000$        &    $2481$      &        $2429$      &    $2264.09$       &     $1.4815$       &   $1\times 10^{-5}$                  \\\hline
\end{tabular}
\end{center}
\end{table}

\begin{table}[h]
\caption{\label{tab:4} }%(absolute value)
\begin{center}
\begin{tabular}{|c|c|c|c|c|c|c|}
\hline
\multicolumn{1}{|c|}{Graph}  & \multicolumn{1}{c|}{$|V|$} & \multicolumn{1}{c|}{$f_{prev}$  } & \multicolumn{1}{c|}{$f_{best}$} & \multicolumn{1}{c|}{$f_{avg}$} & \multicolumn{1}{c|}{$t(s)$}   & \multicolumn{1}{c|}{$P$} \\ \cline{1-7} \hline
$G_{43}$ &       $1000$        &    $6660$      &        $6659$      &    $6602.99$       &     $1.3895$         & $9\times 10^{-5}$                 \\
$G_{44}$     &   $1000$       & $6650$      &        $\textbf{6650}$    &  $6598.03$  &    $1.4145$         & $1\times 10^{-5}$            \\
$G_{45}$   &     $1000$      &   $6654$     &    $6652$              & $6596.63$       &         $1.4245$          &$13\times 10^{-5}$            \\ 
$G_{46}$ &        $1000$         &    $6649$      &         $\textbf{6649}$      &    $6596.22$       &     $1.448$      & $3\times 10^{-5}$                   \\
$G_{47}$ &       $1000$       &    $6657$      &        $6656$      &    $6533.41$       &     $0.5085$       & $6\times 10^{-5}$                   \\
$G_{48}$ &        $1000$        &    $6000$      &        $\textbf{6000}$      &    $5962.85$       &     $7.6625$   &  $0.79265$                         \\
$G_{49}$ &       $1000$      & $6000$      &        $\textbf{6000}$    &  $5961.08$  &    $6.1675$            & $0.61441$          \\
$G_{50}$ &     $1000$      &   $5880$     &    $\textbf{5880}$              & $5843.4$       &        $40.2095$ & $0.06041$                  \\ 
$G_{51}$ &       $1000$        &    $3848$      &         $3835$      &    $3765.98$       &     $0.5403$        & $10^{-6}$                   \\
$G_{52}$ &       $1000$       &    $3851$      &        $3836$      &    $3769.69$       &     $0.5344$      & $10^{-6}$                       \\
$G_{53}$ &      $1000$       &    $3850$      &        $3839$      &    $3767.53$       &    $0.5445$         & $10^{-6}$                    \\
$G_{54}$ &      $1000$       &   $3852$     &    $3832$              & $3767.07$       &         $0.5424$    & $10^{-6}$                  \\ 
$G_{55}$ &       $5000$        &    $10294$      &         $10248$      &    $9958.6$       &     $7.7002$    & $10^{-6}$                    \\
$G_{56}$ &       $5000$        &    $4012$      &        $3977$      &    $3681.49$       &     $7.7121$      & $10^{-6}$                        \\
$G_{57}$ &       $5000$        &    $3492$      &        $3456$      &    $3270.02$       &     $6.7078$      & $10^{-6}$                    \\
%$G_{58}$ &     $5000$       &   $19263$     &    $$              & $$       &         $$                 \\ 
$G_{59}$ &       $5000$        &    $6078$      &         $5969 $      &    $5836.16
$       &     $14.0284$    & $3\times 10^{-5}$                    \\
$G_{60}$ &       $7000$        &    $14176$      &        $14113 $      &    $13987.8$       &     $19.3636$      & $3\times 10^{-5}$                       \\
$G_{61}$ &       $7000$        &    $5789$      &        $5711$      &    $5588.15$       &     $19.3446$   & $3\times 10^{-5}$                         \\
$G_{62}$ &       $7000$        &    $4868$      &        $4810 $      &    $4671.66
$       &     $13.0242$   & $3\times 10^{-5}$                          \\
$G_{63}$ &       $7000$        &    $26997$      &        $26819 $      &    $26728.2$       &     25.0286$$       & $3\times 10^{-5}$                    \\
$G_{64}$ &       $7000$        &    $8735$      &        $8575$      &    $8382.28
$       &     $19.881$      & $3\times 10^{-5}$                      \\
$G_{65}$ &       $8000$        &    $5558$      &        $5494$      &    $5330.1$       &     $9.3802$                        & $3\times 10^{-5}$   \\
$G_{66}$ &       $9000$        &    $6360$      &        $6250$      &    $6040.66$       &     $10.76$    & $3\times 10^{-5}$                       \\
$G_{67}$ &       $10000 $        &    $6940$      &        $6860 $      &    $6040.66$       &     $12.78$      & $1\times 10^{-4}$                  \\
$G_{70}$ &       $10000 $        &    $9541$      &        $9511 $      &    $9363.93$       &     $5.75$       & $6\times 10^{-6}$                  \\
$G_{72}$ &     $10000 $       &   $6998$     &    $6886 $              & $6710.08$       &         $13.95$     & $3\times 10^{-5}$              \\ 
$G_{77}$ &     $14000$       &   $9926$     &    $9788 $              & $9539.19$       &         $26.62$    & $3\times 10^{-5}$               \\
$G_{81}$ &     $20000 $       &   $14030$     &    $ 13818$              & $13476.3$       &         $30.27$     & $3\times 10^{-5}$             \\  \hline
\end{tabular}
\end{center}
\end{table}

\begin{table}[h]
\caption{\label{tab:3_1} Computational results for SA, NMFA, SimCIM on the matrix $N=500$. In columns the Ising energy (absolute and mean values) and mean runtimes (in minutes) are shown.}
\begin{center}
\begin{tabular}{|c|c|c|c|c|c|c|c|}
\hline
\multirow{3}{*}{Algorithm}  & \multicolumn{6}{c|}{Amount of MC steps }                                                                                               \\ \cline{2-7} 
                        & \multicolumn{2}{c|}{$10^3$} & \multicolumn{2}{c|}{$10^4$} & \multicolumn{2}{c|}{$10^5$}            \\ \cline{2-7} 
                        & Best (Mean)    & Time   & Best (Mean)     & Time    & Best (Mean)      & Time              \\ \hline
SA           &        $3776.61$      &  $0.00450822$      &   $4694.47$        &   $0.00767397$      &                  $4861.87$      & $0.428646$                   \\ 
          &     $(3294.65)$           &        &      $(3419.85)$            &         &          $(4712.23)$          &                                         \\ \hline
NMFA      &   $4852.66$  &   $ 0.0177885$     &    $4861.14$   & $0.0954856$     &   $4858.63$        &  $1.13219$                      \\ 
         &     $(4721.6)$          &        &   $(4735.71)$              &         &     $(4821.82)$             &           \\ \hline
SimCIM &         $4859.53$      &  $0.0331745$   &    $4865.16$  &    $0.267649$ &  $4863.14$  &     $ 2.62688$        \\ 
           &      $(4700.7)$     &        &         $(4838.58)$          &         &     $(4861.12)$             &             \\ \hline
\end{tabular}
\end{center}
\end{table}
\begin{table}[h]
\caption{\label{tab:4_1}  Computational results for SA, NMFA, SimCIM on the matrix $N=1000$. In columns the Ising energy (absolute and mean values) and mean runtimes (in minutes) are shown.}
\begin{center}
\begin{tabular}{|c|c|c|c|c|c|c|c|}
\hline
\multirow{3}{*}{Algorithm}  & \multicolumn{6}{c|}{Amount of MC steps }                                                                                               \\ \cline{2-7} 
                        & \multicolumn{2}{c|}{$10^3$} & \multicolumn{2}{c|}{$10^4$} & \multicolumn{2}{c|}{$10^5$} \\ \cline{2-7} 
                        & Best (Mean)    & Time   & Best (Mean)     & Time    & Best (Mean)      & Time               \\ \hline
SA           &      $7927.8$        &   $0.0321744$    &      $12844.7$    &    $0.267792 $    &               $13709.8$       &   $ 3.03297$                          \\ 
          &     $(6705.54)$           &        &              $(12222.1)$    &         &        $(13461.9)$             &                              \\ \hline
NMFA      &   $13642.4$  &   $0.0714152$     &    $ 13743.9$   & $0.649936$     &   $13657.7$        &  $6.43747$          \\ 
         &     $(13436.4)$          &        &   $(13604.8)$              &         &     $(13598.7)$             &                       \\ \hline
SimCIM &       $13799.2$      &  $0.097289$   &    $13808.$  &    $0.941606$ &  $13751.6$  &     $8.41967$           \\ 
           &      $(13591.5)$     &        &         $(13799.5)$          &         &     $(13751.6)$             &                      \\ \hline
\end{tabular}
\end{center}
\end{table}
\begin{table}[h]
\caption{\label{tab:5}  Computational results for SA, NMFA, SimCIM on the matrix $N=2000$. In columns the Ising energy (absolute and mean values) and mean runtimes (in minutes)are shown.}
\begin{center}
\begin{tabular}{|c|c|c|c|c|c|c|c|}
\hline
\multirow{3}{*}{Algorithm}  & \multicolumn{6}{c|}{Amount of MC steps }                                                                                               \\ \cline{2-7} 
                        & \multicolumn{2}{c|}{$10^3$} & \multicolumn{2}{c|}{$10^4$} & \multicolumn{2}{c|}{$10^5$} \\ \cline{2-7} 
                        & Best (Mean)    & Time   & Best (Mean)     & Time    & Best (Mean)      & Time               \\ \hline
SA           &      $15869.3$        &   $ 0.12409$    &      $33281.4$    &    $ 0.956038$    &               $37911. $       &   $9.98077$                          \\ 
          &     $(13103.8)$           &        &              $(31917.3)$    &         &        $(37326.8)$             &                              \\ \hline
NMFA      &   $38604.3$  &   $0.22096$     &    $38806.2$   & $1.99608$     &   $38690.6$        &  $19.2767$          \\ 
         &     $(37865.3)$          &        &   $(38460.7)$              &         &     $(38537.9)$             &                       \\ \hline
SimCIM &       $39063.3$      &  $0.262565$   &    $39039.7$  &    $ 2.58162$ &  $38903.5 $  &     $26.7321$           \\ 
           &      $(38628.7)$     &        &         $(38857.3)$          &         &     $(38903.)$             &                      \\ \hline
\end{tabular}
\end{center}
\end{table}

 \begin{table}[h]
\caption{\label{tab:3_2} Computational results for MARS on the matrix $N=\{500,1000,2000\}$. In columns the Ising energy (absolute and mean values) and exeptence probability are shown.}
\begin{center}
\begin{tabular}{|c|c|c|c|c|c|c|c|}
\hline
\multirow{3}{*}{N}  & \multicolumn{6}{c|}{Amount of simulation }                                                                                               \\ \cline{2-7} 
                        & \multicolumn{2}{c|}{$10^2$} & \multicolumn{2}{c|}{$10^3$} & \multicolumn{2}{c|}{$10^4$}            \\ \cline{2-7} 
                        & Best (Mean)    & P   & Best (Mean)     & P   & Best (Mean)      & P              \\ \hline
           $500$&         $4865.16$      &  $0.09$   &    $4865.16$  &    $0.113$ &  $4865.16$  &     $0.0099$        \\ 
           &      $(4762.55)$     &        &         $(4715.56)$          &         &     $(4713.3)$             &             \\ \hline
                      $1000$&         $13784.4$      &  $0.0202$   &    $13822.1$  &    $0.001$ &  $13822.1$  &     $0.0002$        \\ 
           &      $(13470.1)$     &        &         $(13464.6)$          &         &     $(13463.5)$             &             \\ \hline
             $2000$&         $39037.2$      &  $0.01$   &    $39091.1$  &    $0.001$ &  $39070.6$  &     $0.0002$        \\ 
           &      $(38471)$     &        &         $(38115.7)$          &         &     $(38108.7)$             &             \\ \hline
\end{tabular}
\end{center}
\end{table}
 \begin{table}[h]
\caption{\label{tab:3_3} Computational results for MARS on the matrix $N=\{500,1000,2000\}$ for the different temperature boudaries. In columns the Ising energy (absolute and mean values) and exeptence probability are shown.}
\begin{center}
\begin{tabular}{|c|c|c|c|c|c|c|c|}
\hline
\multirow{3}{*}{T}  & \multicolumn{6}{c|}{Amount of simulation }                                                                                               \\ \cline{2-7} 
                        & \multicolumn{2}{c|}{$500$} & \multicolumn{2}{c|}{$1000$} & \multicolumn{2}{c|}{$2000$}            \\ \cline{2-7} 
                        & Best (Mean)    & P   & Best (Mean)     & P   & Best (Mean)      & P              \\ \hline
           $[0,10]$&         $4865.16$      &  $0.0009$   &    $13778.2$  &    $0.0001$ &  $38595.2$  &     $0.0001$        \\ 
           &      $(4610.75)$     &        &         $(13028.7)$          &         &     $(36719.1)$             &             \\ \hline
                      $[10,20]$&         $4865.16$      &  $0.1948$   &    $13822.1$  &    $0.0003$ &  $39051.8$  &     $0.0001$        \\ 
           &      $(4816.23)$     &        &         $(13604)$          &         &     $(38125.5)$             &             \\ \hline
             $[20,30]$&         $-$      &  $-$   &    $13798.1$  &    $0.0006$ &  $39091.1$  &     $0.0002$        \\ 
           &      $(-)$     &        &         $(13749.6)$          &         &     $(38711.9)$             &             \\ \hline
\end{tabular}
\end{center}
\end{table}

%\end{thebibliography}
\end{document}